\documentclass{jac}
\usepackage{graphicx}

\newcommand{\ZZ}{\mathbb{Z}}
\newcommand{\NN}{\mathbb{N}}
\newcommand{\C}{\mathfrak{C}}
\newcommand{\PP}{\mathcal{P}}
\newcommand{\F}{\mathcal{F}}
\begin{document}
    
\title{Yet Another Aperiodic Tile Set}
\author
{V.~Poupet}{Victor Poupet}
\thanks{The author is partly supported by ANR-09-BLAN-0164.}
\address
{LIF Marseille}
\email{victor.poupet@lif.univ-mrs.fr}


\begin{abstract}
    \noindent We present here an elementary construction of an aperiodic tile set. Although there already exist dozens of examples of aperiodic tile sets we believe this construction introduces an approach that is different enough to be interesting and that the whole construction and the proof of aperiodicity are hopefully simpler than most existing techniques.
\end{abstract}

\maketitle

Aperiodic tile sets have been widely studied since their introduction in 1962 by Hao Wang \cite{KahrMooreWang62}. It was initially conjectured by Wang that it was impossible to enforce the aperiodicity of a coloring of the discrete plane $\ZZ^2$ with a finite set of local constraints (there either was a valid periodic coloring or none at all). This would imply that it was decidable whether there existed a valid coloring of the plane for a given set of local rules. This last problem was introduced as the \emph{domino problem}, and eventually proved undecidable by Robert Berger in 1964 \cite{Berger64, Berger66}. In doing so, Berger produced the first known aperiodic tile set: a set of local rules that admitted valid colorings of the plane, none of them periodic. Berger's proof was later made significantly simpler by Raphael M. Robinson in 1971 \cite{Robinson71} who created the set of Wang tiles now commonly known as the \emph{Robinson tile set}.

Since then many other aperiodic tile sets have been found, not only on the discrete plane \cite{DurandLevinShen04, DurandShenRomashchenko08, Kari07, Ollinger08}, but also on the continuous plane \cite{Goodman-Strauss99, Penrose78}.

In this article we will describe yet another construction of an aperiodic tile set. Although the resulting tile set will produce tilings quite similar to those of Robinson's (an infinite hierarchical structure of embedded squares) the local constraints will be presented in a (hopefully) more natural way: we will start from simple geometrical figures and organize them step by step by adding new rules progressively.

\section{Tilings} 
\label{sec:tilings}

There are many different ways to define tilings of the discrete plane $\ZZ^2$. The historical definition as presented by Wang is that of unit square tiles with colored edges (nowadays called \emph{Wang tiles}). The domino problem is to decide whether one can arrange copies of a given set of such tiles on the plane so that the adjacent sides of two neighbor tiles have the same color.

Although the description of the problem with Wang tiles is extremely simple, it is not the easiest  way to deal with tilings. A more modern approach consists in defining a tile set as a set of local constraints in the form of a finite set of forbidden patterns. A tiling of the plane according to such a tile set is a coloring of the plane such that no forbidden pattern appears. Both definitions are known to be equivalent.

\subsection{Patterns and Configurations}

\begin{definition}[Configuration]\label{def:configuration}
	Given a finite set of symbols $\Sigma$, a $\Sigma$-\emph{configuration} is a mapping $\C:\ZZ^2\rightarrow \Sigma$ that associates a symbol of $\Sigma$ to each element of $\ZZ^2$ (elements of the plane $\ZZ^2$ will be referred to as \emph{cells}).
\end{definition}

\begin{definition}[Pattern]\label{def:pattern}
	Given a finite set of symbols $\Sigma$, a $\Sigma$-\emph{pattern} is a mapping $\PP: D_\PP \rightarrow \Sigma$ from a finite subset of cells $D_\PP\subset \ZZ^2$ to $\Sigma$.
\end{definition}

\begin{definition}[Tile Set]\label{def:tile_set}
	A \emph{tile set} is a couple $\tau=(\Sigma, \F)$ where $\Sigma$ is a finite set of symbols and $\F$ is a finite set of patterns called \emph{forbidden patterns}.
\end{definition}

\begin{definition}[Tilings]\label{def:valid}
	Given a finite set of symbols $\Sigma$, we say that a $\Sigma$-pattern $\PP:D_\PP\rightarrow \Sigma$ \emph{appears} in a $\Sigma$-configuration $\C$ if there exists a vector $v\in\ZZ^2$ such that
	\[
	\forall x \in D_\PP, \PP(x) = \C(x+v)
	\]
	
	 A $\Sigma$-configuration $\C$ is said to be \emph{valid} for a tile set $\tau=(\Sigma, \F)$ if it contains none of the patterns in $\F$. A tiling of the plane by a tile set $\tau$ is a valid configuration for $\tau$.
\end{definition}

\subsection{Periodicity}

\begin{definition}[Periodicity]\label{def:periodicity}
	A configuration $\C$ is said to be \emph{periodic} if there exists a non-zero vector $v\in\ZZ^2$ such that $\C$ is invariant by a translation of $v$ ($v$ is a vector of periodicity of $\C$):
	\[\forall x\in \ZZ^2, \C(x) = \C(x+v)\]
	A configuration is said to be \emph{bi-periodic} if it has two independent vectors of periodicity.
\end{definition}

\begin{definition}[Aperiodicity]\label{def:aperiodicity}
	A tile set is said to be \emph{aperiodic} if it admits at least one tiling of the plane but admits no periodic tiling.
\end{definition}
The two following propositions will be of use later. The first one is folklore and its proof will be omitted.

\begin{proposition}
	\label{pro:bi-periodicity}
    If a tile set admits a valid periodic tiling of the plane it admits a valid bi-periodic tiling of the plane.
\end{proposition}

\remark The contraposition of Proposition~\ref{pro:bi-periodicity} states that if a tile set cannot tile the plane bi-periodically it cannot tile it periodically either. We will use this in our construction of an aperiodic tile set as it is easier to prove that there exist no bi-periodic valid configuration.

\begin{proposition}\label{pro:vertical}
	If a configuration is bi-periodic it has both a vertical and a horizontal vector of periodicity.
\end{proposition}
\proof
If $(x,y)$ and $(x', y')$ are two independent vectors of periodicity of a configuration then $x'.(x, y) - x.(x',y') = (0, x'y - xy')$ and $y'.(x,y)-y.(x',y')=(xy'-x'y, 0)$ also are.
\qed

\section{Construction of an Aperiodic Tile Set} 
\label{sec:the_main_construction}

\subsection{General Overview} 
\label{sub:general_overview}

The aperiodic tile set that we are going to describe is based on the following simple observation: if a picture contains arbitrarily large squares such that none of these squares intersect each other, the picture cannot be periodic. Indeed, because squares do not intersect a translation vector that leaves the picture unchanged must be larger than the side of every square for if a square is translated less than the length of its sides it intersects its original position.

What we will do now is design a set of local constraints that only accepts pictures on the discrete plane that contain arbitrarily large non-intersecting squares.

These ``pictures'' will contain lines of different sorts made of horizontal, vertical and diagonal segments. To be consistent with the previous definitions of configurations, tile sets and tilings we should be describing configurations as symbols on the cells of $\ZZ^2$. However it will be much easier to explain (and understand) the construction by describing geometrical shapes.

This means that when we will say something like ``blue lines are made of horizontal and vertical segments, have no extremities and cannot cross'' what this really means is that we have symbols representing blue lines going through cells vertically, horizontally and changing directions (for example entering from the top side and exiting from the right side). Once represented, it is easy to enforce the stated properties with a set of forbidden patterns. In this case the forbidden patterns are those where a blue line is interrupted because it exits a cell from one side but does not enter its neighbor from the corresponding side. The fact that blue lines cannot cross is simply enforced by having no symbol corresponding to a crossing on a cell (no symbol corresponds to a blue line going through a cell both vertically and horizontally).

Our construction will consist in two main types of lines that we will call ``blue lines'' and ``arms''. Blue lines will be made to draw non intersecting squares while arms will be used to control the size of squares and connect them together to build a structure that enables us to prove the existence of arbitrarily large squares.


\subsection{Blue Lines} 
\label{sub:blue_lines}

Blue lines are made of vertical and horizontal segments. They have no extremities (only infinite or closed paths) and cannot cross or overlap. They are oriented (they have an \emph{inner} and an \emph{outer} side) and can only change their direction by turning towards the inside.

All of these rules are local conditions and can therefore be enforced by a tile set.

\begin{figure}[htbp]
	\includegraphics[width=6cm]{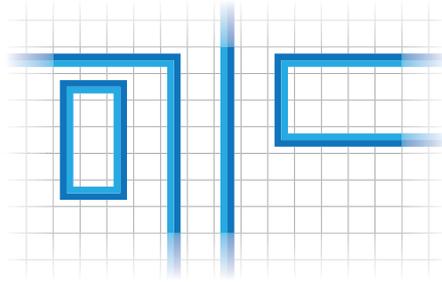}
	\caption{Possible blue paths. The infinite paths can have 0, 1 or 2 angles.}
	\label{fig:blue_rectangles}
\end{figure}

Because blue lines cannot cross and can only turn towards the inside, finite blue lines can only be rectangles. For the same reasons, infinite blue lines can only be of three kinds, each corresponding to a degenerate rectangle with some bi-infinite or semi-infinite sides (see Figure \ref{fig:blue_rectangles}).

\subsection{Blue Squares} 
\label{sub:blue_squares}

We now want to make sure that only squares are valid. To do so, the usual method is to draw a diagonal line from the upper-left and lower-right angles of every blue rectangle. This diagonal line is oriented towards the inside of the rectangle and is not allowed to meet a blue line other than the angles from which it starts. If the rectangle is a square, the two diagonal lines merge into one but if it is not a square the diagonals will reach a side of the rectangle, which is forbidden.

\begin{figure}[htbp]
	\includegraphics[height=7cm]{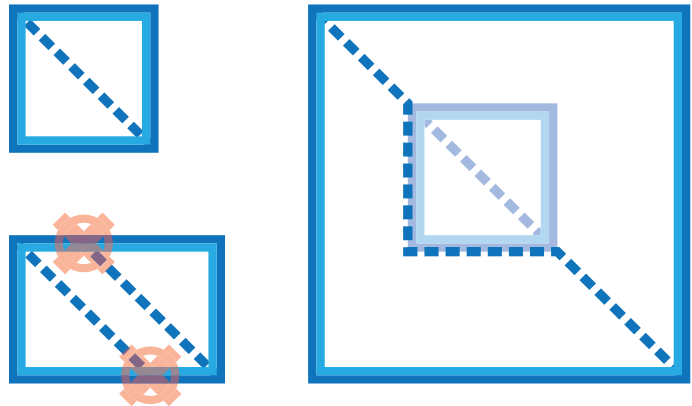}
	\caption{Diagonal line used to ensure all closed blue paths are squares}
	\label{fig:diagonals}
\end{figure}

This however only works if there are no smaller squares inside larger ones. Because we need some small blue squares to lie on the diagonal of larger ones, we will have to allow the diagonal line to ``go around'' a square, but only from one corner to the other as shown in Figure~\ref{fig:diagonals}.

We can show inductively that all closed blue paths are squares:
\begin{itemize}
    \item if a path has no other blue path inside, the diagonal line goes straight from its upper-left to its lower-right corners, it is therefore a square;
    \item if all blue paths inside a larger one are squares, the diagonal line only goes around squares and hence it remains on the real diagonal of the larger one, the large one is a square too.
\end{itemize}

\remark A small blue square inside a larger one can only be either perfectly aligned with the latter's diagonal or far enough from it so that it does not intersect it.

\subsection{Infinite Paths} 
\label{sub:infinite_paths}

\begin{lemma}\label{lem:infinite}
    The only possible infinite blue paths in a valid bi-periodic configuration are infinite straight lines (no angle).
\end{lemma}
\proof
According to the basic rules of blue lines, infinite blue paths can be of three different kinds (illustrated by Figure \ref{fig:blue_rectangles}): they can have zero, one or two angles.

\begin{figure}[htbp]
	\includegraphics[height=5cm]{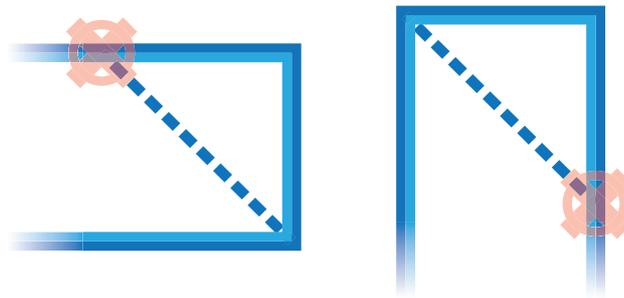}
    \caption{Two-angled infinite paths cannot be valid.}
    \label{fig:two_angles}
\end{figure}

However, because of the diagonal line that starts from the upper left and lower right angles of any blue line, infinite paths with two angles cannot be valid (see Figure \ref{fig:two_angles}).

Moreover, no valid bi-periodic configuration can contain an infinite path having only one angle. Indeed, such a configuration must be both horizontally and vertically periodic (Proposition~\ref{pro:bi-periodicity}) and any finite horizontal or vertical translation of the infinite angle would intersect it. 

The only remaining case of infinite blue path is that of bi-infinite vertical or horizontal straight lines (with no angle).
\qed

\subsection{Arms} 
\label{sub:arms}

\begin{figure}[htbp]
    \centering
    \includegraphics[height=7cm]{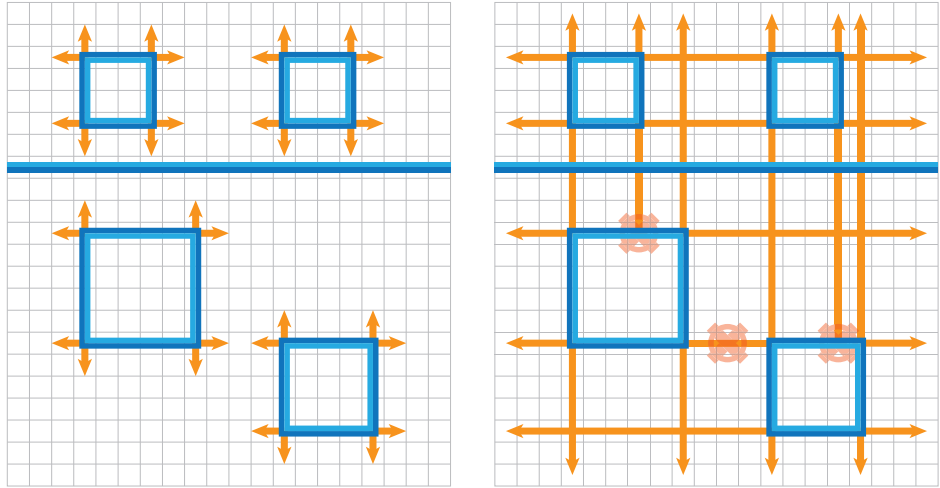}
    \caption{Arms extend from every angle of a blue path towards the exterior. In the right part there are three errors: the middle one is an orientation error (down side connected to an up side) while the two others are arms that cross more than one blue line.}
    \label{fig:arms}
\end{figure}

Blue squares alone are not sufficient to ensure the aperiodicity of the tile set. What we will do now is organize them into groups in such a way that for every blue square of finite size we can prove the existence of a larger finite blue square. In order to group them, we extend vertical and horizontal lines from every corner of a blue square towards the exterior (see Figure~\ref{fig:arms}). These new lines are called \emph{arms}.

The basic properties of arms can be described by the following rules:
\begin{itemize}
    \item Arms are horizontal or vertical continuous straight lines. They do not turn.
    \item Arms and blue lines cannot overlap.
    \item Arms are allowed to cross other perpendicular arms.
    \item The extremities of an arm must be angles of blue paths (some extremities might not exist if the arm is semi or bi-infinite).
    \item The orientations of two blue squares connected by an arm must match: an arm cannot connect the upper (resp. right) side of a square to the lower (resp. left) side of another.
    \item There can be at most one point on an arm where it crosses a blue line.
\end{itemize}

\begin{figure}[htbp]
    \centering
    \includegraphics[height=4cm]{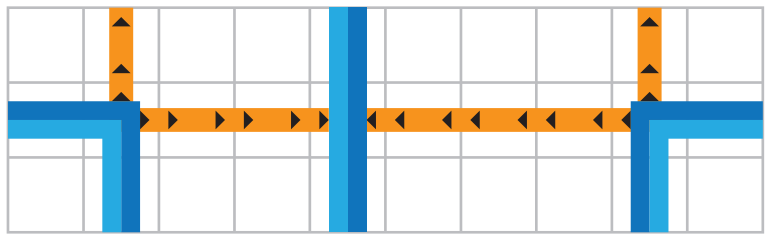}
    \caption{Orientation on the arms to enforce locally the fact that an arm can cross at most one blue line. The blue line can only cross where the orientations meet.}
    \label{fig:arms_oriented}
\end{figure}

The last rule is the key to most of the properties that we will need later. It might appear as a non-local constraint as it is formulated as a global condition on the arm but it can be enforced locally by orienting the arms from their extremities as shown in Figure \ref{fig:arms_oriented}: blue lines are only allowed to cross an arm where the two opposite orientations meet (which needs not be the middle of the arm).

Two blue squares are said to be \emph{neighbors} if they are connected by an arm.
\subsection{Size Matching} 
\label{sub:size_matching}

\begin{figure}[htbp]
    \centering
        \includegraphics[width=12cm]{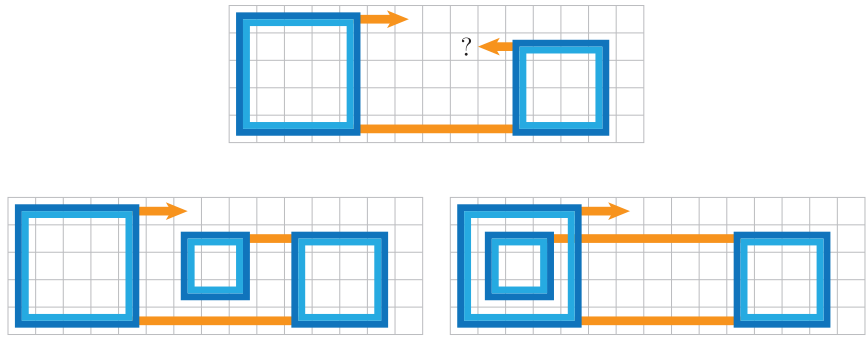}
    \caption{Why arms cannot connect squares of different sizes.}
    \label{fig:size}
\end{figure}

\begin{lemma}\label{lem:size}
    In a valid bi-periodic configuration, if two blue squares are neighbors they are of equal size.
\end{lemma}
\proof
By contradiction, let us assume there exists a valid bi-periodic configuration having two connected squares of different size. Let us consider one of the smallest squares so connected to a larger square. In order to describe the situation, we will consider that the two squares are connected horizontally by an arm joining their lower sides (as shown in Figure~\ref{fig:size}).

There has to be an arm that starts from the upper side of the smaller square and goes towards the larger. Because this arm is not allowed to cross two blue lines, it cannot go entirely through the larger square. Thus it must be connected to the upper side of another blue square, either before entering the larger square or inside it (both cases are illustrated by Figure~\ref{fig:size}). In both cases, this third square at the other extremity of the arm must be smaller than the initial small square~:
\begin{itemize}
    \item if the third square is outside of the larger one, its vertical sides cannot cross the arm connecting the two initially considered squares for this would mean this arm is crossed by two blue lines;
    \item if the third square is inside the larger one, it cannot cross the side of the larger square;
\end{itemize}

This contradicts the fact that the initial square was chosen as being one of the smallest squares connected to a square of different size.
\qed

\begin{lemma}\label{lem:neighbors}
    In a valid bi-periodic configuration, every finite blue square has exactly four neighbors, one in each direction, and it is connected to each of its neighbors by two arms.
\end{lemma}
\proof Because all connected squares have the same size, if two squares are connected by an arm, they are also connected by a second arm. Since every finite blue square has eight arms, it is connected to at most four neighbors.

Moreover there can be no semi-infinite horizontal or vertical line in a configuration that is both vertically and horizontally periodic (the line would have to be bi-infinite) hence every arm connected to a square is connected to another one. Every square must then have at least four neighbors, one in each direction.
\qed

\subsection{Groups} 
\label{sub:groups}
It is now time to add a communication between the different finite blue squares in order to organize them in groups in such a way that for each group of neighbor squares there exists a larger finite square associated with this group, in turn leading to the proof that there exist arbitrarily large finite squares.

To do this, we add two coordinates $(x, y)\in (\ZZ/3\ZZ)^2$ to every blue line, and the arms only allow a connection that corresponds to a correct arrangement of squares: the right neighbor of a square $(x, y)$ must have coordinates $(x +1, y)$ and its up neighbor must have coordinates $(x, y+1)$ (all additions are performed modulo $3$).

To realize this with a tile set we use different sorts of blue lines for each possible set of coordinates (9 possibilities), and different sorts of arms depending on the coordinates of the squares they connect. Obviously we require that the coordinates of a blue line are constant along the line (which is a local condition) and that the coordinates of an arm (the coordinates of the squares it connects) are also constant along one arm.

The structure is then enforced by the arms at their extremities: a horizontal arm whose coordinates are $((x, y), (x+1, y))$ must be connected to a square of coordinates $(x,y)$ by its left extremity and to a square of coordinates $(x+1, y)$ by its right extremity, and similarly with vertical arms of coordinates $((x, y), (x, y+1))$.

\begin{lemma}\label{lem:oneone}
In a valid bi-periodic configuration, if there exists a blue square then there exists a blue square of the same size with coordinates $(1, 1)$.
\end{lemma}
\proof This is a straightforward consequence of Lemmas \ref{lem:size} and \ref{lem:neighbors} and the coordinates system. All squares have neighbors in all directions and all neighbors have the same size. Because the first (resp. second) coordinate is incremented by $1$ modulo $3$ each time we consider the right (resp. up) neighbor, we eventually find a square of coordinates $(1, 1)$.
\qed

The construction is now nearing its end. All we need to do is ensure that in any possible bi-periodic configuration, for every finite blue square there exists another blue square that is larger. The easy way to prove that a square is larger than another is to have the large one contain the other. Because for every square there is a $(1, 1)$ square of the same size, it is enough to make every $(1, 1)$ square be inside a larger one.

\begin{figure}[htbp]
    \centering
        \includegraphics[width = 10cm]{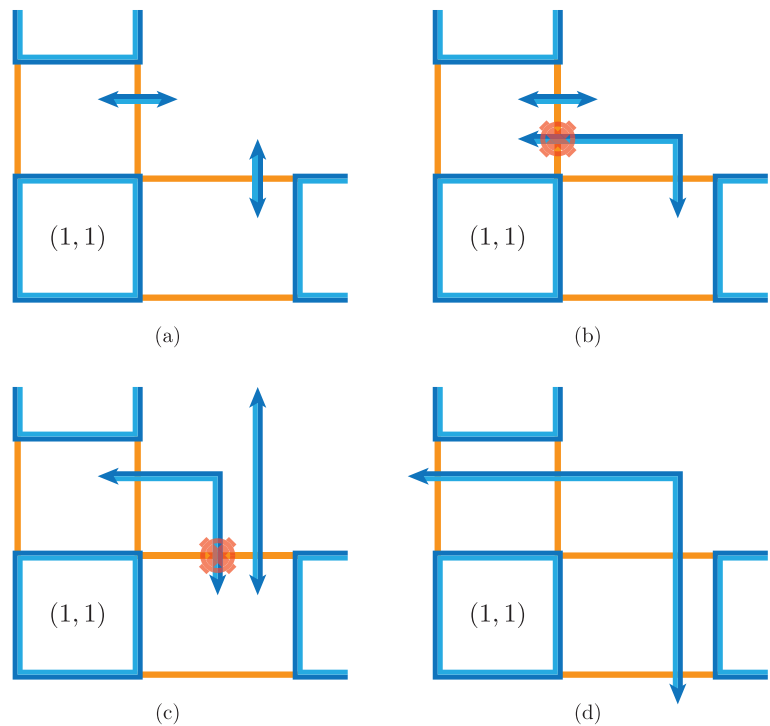}
    \caption{The two blue lines that cross the right and top arms of a $(1,1)$ square are connected.}
    \label{fig:large_squares}
\end{figure}

To do so, we slightly change the arms connecting $(1, 1)$ squares to their neighbors. Instead of being allowed to cross at most one blue line, these arms are \emph{required} to cross exactly one blue line. Moreover the inner side of the crossing blue line must be towards the $(1,1)$ square. This is easy to do with local constraints by requiring a blue line to cross such an arm where the opposite orientations meet (as explained in sub-section \ref{sub:arms} and illustrated by Figure \ref{fig:arms_oriented}). We can now prove the following lemma:

\begin{lemma}\label{lem:large_squares}
    In a valid bi-periodic configuration, every finite $(1,1)$ blue square is contained in a larger finite blue square.
\end{lemma}
\proof
Consider a finite $(1, 1)$ blue square. By lemma \ref{lem:neighbors} it has both an up and a right neighbor. The arms that connect it to these neighbors are each crossed by a blue line, with its inside turned towards the $(1,1)$ square. The situation is illustrated in Figure \ref{fig:large_squares} (a). The two blue lines that cross the arms must be connected:
\begin{itemize}
    \item if the vertical one turns before the position of the horizontal one, it will have to cross the arm that is already crossed by the horizontal portion of blue line (b) or turn once more and move a second time though the arm it has already crossed once;
    \item if the vertical blue line goes further up than the position of the horizontal one, the horizontal one must turn before and cross one of the two arms that have already been crossed (c).
\end{itemize}
The two blue lines that cross the arms are two sides of the same blue path (d). This blue path contains the $(1,1)$ square and is therefore larger.
\qed
\subsection{Aperiodicity} 
\label{sub:aperiodicity}
All we need to do now is make sure there is a blue square somewhere in any valid configuration. The simplest way to do this locally is to forbid large patterns that have no blue angle. In our specific case, patterns of size 2 are sufficient so we add this last rule to our tiling constraints: every $2\times 2$ pattern must contain a blue angle.

We can now prove the key proposition of the construction:
\begin{proposition}\label{pro:aperiodicity}
    There exists no valid periodic configuration.
\end{proposition}
\proof
By Proposition \ref{pro:bi-periodicity}, we need only show that there exists no valid bi-periodic configuration. As a consequence of the last rule any valid configuration has a blue angle. By Lemma \ref{lem:infinite} this angle is part of a finite blue square. Finally, by Lemmas \ref{lem:oneone} and \ref{lem:large_squares} for every finite blue square in a valid bi-periodic configuration there exists a larger finite blue square. This means that any such configuration contains arbitrarily large non-intersecting blue squares, which contradicts its periodicity.
\qed

\subsection{Valid Configuration} 
\label{sub:valid_configuration}
\begin{figure}[htbp]
	\centering
	\includegraphics[width = 15cm]{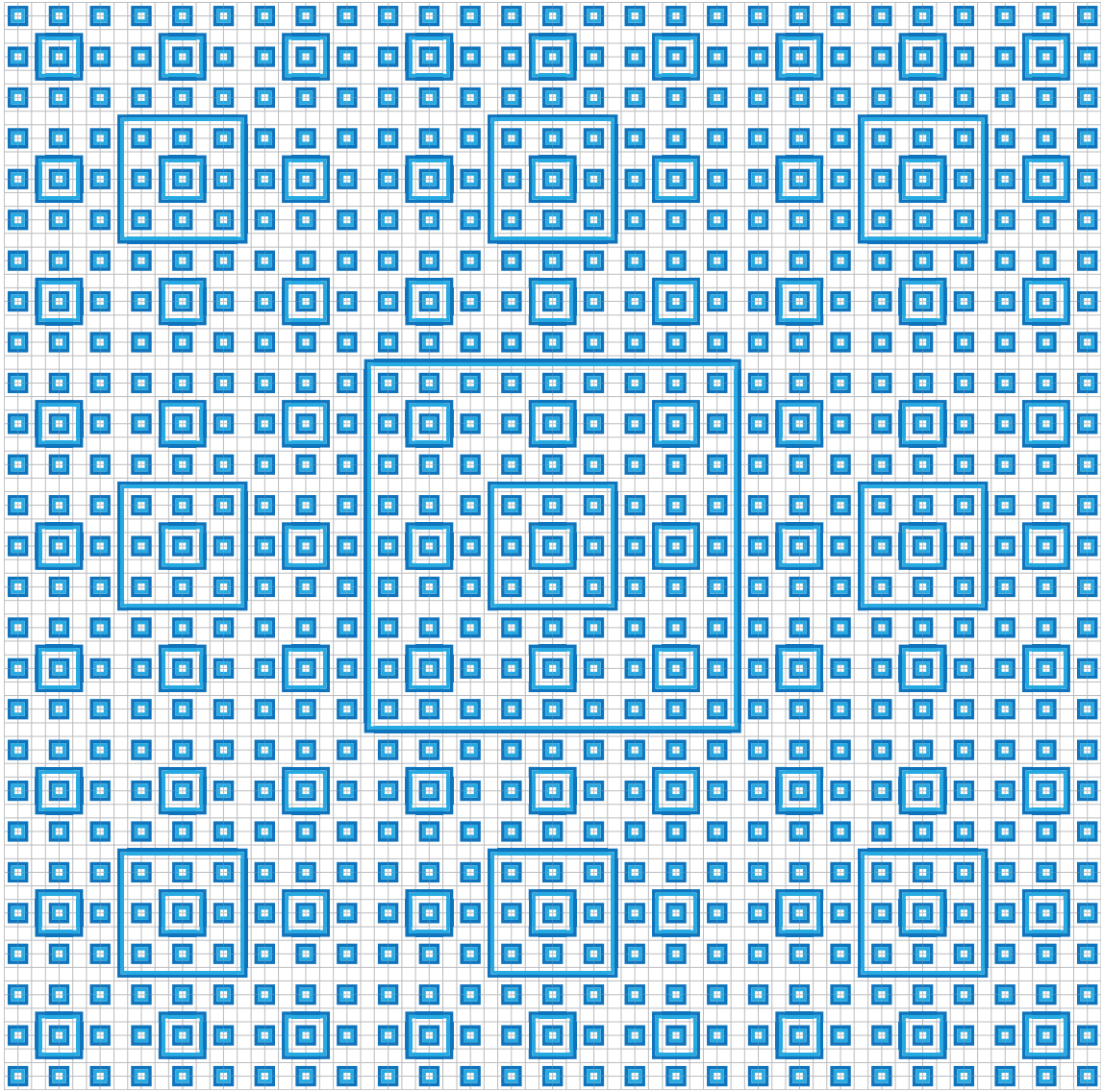}
	\caption{A valid configuration}
	\label{fig:configuration}
\end{figure}

We still need to show that there exists at least one valid configuration, for the tile set would otherwise be of very limited interest. We will now show that the configuration illustrated by Figure~\ref{fig:configuration} is valid.

This configuration is very regular and has a simple structure. It contains squares of size $3^k$ for every $k\in\NN$. For every $k$, the squares of size $3^k$ are arranged regularly, each being at a distance $2.3^k$ from its neighbors. They are then considered in groups of $3\times 3$ and there is a square of size $3^{k+1}$ that has the same center as the central square of size $3^k$ of each group. The central square in each $3\times 3$ group has coordinates $(1,1)$ while the others have the matching coordinates (the arms are not represented in the Figure for better clarity).

This configuration satisfies all the rules of the tile set:
\begin{itemize}
	\item blue paths are all finite blue squares;
	\item because squares are so regularly arranged, smaller squares that intersect the diagonal of a larger one are perfectly aligned with this diagonal;
	\item arms connect squares of the same size, and every square has four neighbors;
	\item the squares of size $3^{k+1}$ contain the squares of size $3^k$ that have coordinates $(1,1)$ but none of their neighbors so they cross all the required arms.
\end{itemize}

Two things must still be justified. The first is that the arrangement of squares that has been described can fill an infinite configuration and more precisely that there is always room for the larger squares without overlapping the previously existing lines. This can be proved by observing that between two consecutive columns (resp. rows) of squares of size $3^k$, if we ignore all the larger squares, there is an empty column (resp. row) of cells (cells on which there is no blue line). This property can be proved inductively since it is true for the squares of size $1$ and at each step, the squares of size $3^{k+1}$ occupy two out of three empty columns and rows, leaving exactly one empty column or row between neighboring squares. The construction can therefore be continued indefinitely.

Lastly we must verify that in the valid configuration no arm is crossed by more than one blue line. This fact is closely related to the previous point: the sides of squares of length $3^k$ lie on rows and columns that were not crossed by smaller squares (the empty rows and columns previously discussed). No smaller square can therefore cross arms connecting two squares of side $3^k$. Finally, between two neighboring squares there is exactly one empty column or row on which a larger square could have its side and hence at most one blue line can cross an arm.

\section{Conclusion} 
\label{sec:conclusion}

What we have described is a set of local rules (all rules concern neighboring cells and can be described with $2\times 2$ forbidden patterns) that admits infinite valid configurations but none of these are periodic. Although the local rules remain simple and the number of geometric structures used is quite limited (blue lines, arms and diagonals), the number of symbols necessary to represent them on the cells is very large. Because each cell can contain different combinations of lines and that said lines must be different depending on the information they hold (orientation, coordinates in a group of squares, number of blue lines crossed by an arm, etc.) tens of thousands of different symbols are used.

\begin{figure}[htbp]
	\centering
	\includegraphics[width=10cm]{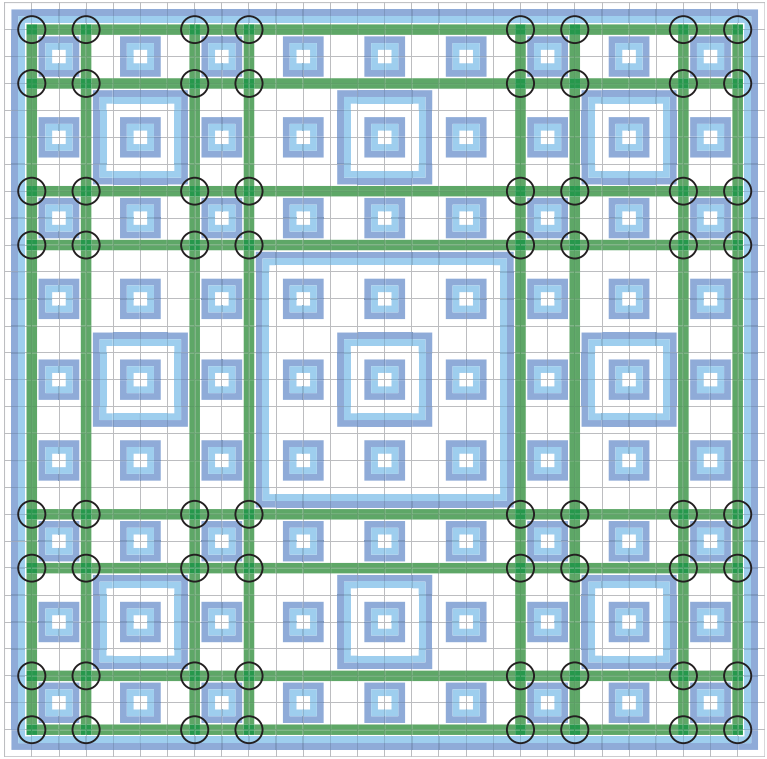}
	\caption{Computation area in a square of size 27. Only the intersections correspond to cells of the space-time diagram of the Turing machine, the horizontal and vertical lines are used to transmit the data. In this example the square can compute $8\times 8$ cells of the space-time diagram.}
	\label{fig:turing}
\end{figure}

In order to keep the construction as simple as possible we have only proved that the tile set was aperiodic but it is not sufficient to prove the undecidability of the domino problem as it is. The construction can be strengthened however by forcing blue squares of length one to be regularly arranged as they are in the configuration described in Subsection \ref{sub:valid_configuration}. It is then possible to show inductively that the larger squares are also regularly arranged by observing the empty columns and rows. By doing so one can then embed partial space-time diagrams of a Turing Machine in the free space of each blue square as it is done in Robinson's construction (see Figure \ref{fig:turing}). If the halting state of the Turing machine is not included in the tile set, large valid space-time diagrams of the Turing machine cannot appear in a tiling. The produced tile set can hence tile the plane if and only if the Turing machine does not halt, which proves the undecidability of the domino problem.

The structure of the valid tilings can also be easily altered. Groups could be larger and their inner structure can be more complex. For instance it would be possible to mimic the behavior of recursive geometric constructions such as the space-filling curves of Peano \cite{Peano1890} or Hilbert \cite{Hilbert1891} to enforce their structure with a tile set.

\bibliographystyle{plain}
\bibliography{JAC}
\end{document}